\documentclass[12pt]{iopart}

\usepackage{tikz}
\usepackage{graphicx}
\usepackage{amssymb}
\usepackage[mathscr]{eucal}
\usepackage[latin1]{inputenc}
\usepackage{url}

\newcommand{\affiliation}[1]{\address{#1}}

\begin{document}

\title{Gif Lectures on direct detection of Dark Matter}

\author{Eric Armengaud\footnote{eric.armengaud@cea.fr}}

\affiliation{Service de Physique des Particules, IRFU,
CEA, F-91191Gif/Yvette Cedex, France  }

\date{20 September 2009}

\begin{abstract}
These notes cover some of the topics associated with direct detection of dark matter at an introductory level. The general principles of dark matter search are summarized. The current status of some experiments is described, with an emphasis on bolometric and noble liquid techniques. Plots and illustrations associated to these notes may be found on transparencies presented during the lecture, on the web site of Gif school 2009~\footnote{Ecole de Gif, Batz sur mer, 21-25 September 2009} (in French) : \url{http://www-subatech.in2p3.fr/gif2009.html}
\end{abstract}

\section{Introduction}\label{sec:introduction}

Current cosmological observations have lead to a concordance model of our Universe, in which a major role is played by a still mysterious dark matter. Dark matter drives the dynamics of galaxies and clusters, and generates the growth of large-scale structures, from initial density perturbations of order $\sim 10^{-5}$ that are measured in the CMB anisotropies, to inhomogeneities of amplitude $\sim$ 1 observed in the large scale structures at low redshift. In the concordance model, dark matter cannot be of baryonic nature since both primordial abundance measurements and CMB anisotropies point towards a small global baryonic density $\Omega_b$.

Dark matter, as well as dark energy, might be a manifestation of the break-down of general relativity on large scales or in the regime of low accelerations. A large litterature is now devoted to possible modifications of the laws of gravitation : in general, these modifications are equivalent to the introduction of new fields in addition to the metric tensor $g_{\mu\nu}$. Another approch is to assume that dark matter is constituted of a gaz of collisionless, stable and massive particles whose nature is still unknown. Indeed several arguments strongly suggest that the Standard Model of particle physics is not complete, at least for energies larger than the electroweak scale. A large number of extensions of the Standard Model provide quite naturally a possible dark matter candidate in their spectrum~\cite{Steffen:2008qp}.

In these notes we present the experimental efforts for direct detection of possible dark matter (DM) particles. All experiments for DM direct detection aim at testing the hypothesis that our own galactic halo is filled with particle dark matter. For several plausible DM hypothesis, like the gravitino in mSUGRA~\cite{Ellis:2003dn}, the interaction rate of DM with ordinary matter is unfortunately far too small to be detected with current technologies. Two main categories of DM particles can currently be probed by direct detection experiments:

\begin{enumerate}
\item Weakly Interacting Massive Particles (WIMPs) which have cross-sections with ordinary matter driven by physics at the electroweak scale. Currently, the search for WIMPs represents by far the strongest effort in direct detection experiments. Several models of new physics at the electroweak scale provide such candidates, like the neutralino in the minimal supersymetric extension of the standard model (MSSM) or the lightest Kaluza-Klein particle (LKP) in universal extra dimension (UED) models. The case for WIMPs is particularly strong due to the so-called ``WIMP miracle": if dark matter particles are thermal relics, then from thermodynamical considerations their current density $\Omega_M \sim 0.3$ implies a typical annihilation cross-section during their freeze-out $\langle \sigma_{a} v \rangle \sim 3\times 10^{-26}$ cm$^3$/s which is characteristic of weak interactions.
\item Axions or axion-like particles are hypothetical pseudo-scalar particles. Axions were initially introduced to solve the CP violation problem in the strong interaction sector through the so-called Peccei-Quinn mechanism, but axion-like particles constitute a generic prediction of some string models. They can constitute (non-thermal) relics of the Big-Bang. The direct search for axion dark matter is possible thanks to their coupling to photons, and will be briefly mentionned in Section~\ref{sec:axions}.
\end{enumerate}

\noindent We now concentrate mostly on the search for WIMPs, which started during the 80s after the seminal works of, eg.~\cite{Goodman:1984dc,Primack:1988zm}. In the most studied models, typical WIMP masses may range from $\sim 10$ GeV to 10 TeV. For kinematic reasons, direct detection experiments aim in general to observe nuclear recoils due to an elastic scattering of WIMPs on the nuclei of a target. The WIMP-nucleon cross section is poorly constrained by current measurements in frameworks like MSSM or UED . It may range typically from $10^{-6}$ to $10^{-12}$ pb~\footnote{1 pb = 1 picobarn = $10^{-36}$ cm$^2$}. The last value corresponds to less than one interaction per ton of detector per year. All the efforts of WIMP experiments are therefore devoted to the ability to detect such a low rate of interactions.

\section{Theory of WIMP direct detection}\label{sec:theory}

The formalism of WIMP direct detection is briefly sketched here. More details may be found in~\cite{Jungman:1995df,Lewin:1995rx} .

\subsection{Interaction rate}

Let us consider a WIMP of mass $m_{\chi}$ diffusing elastically on a nucleus with mass $m_N$. The nucleus recoils with an angle $\theta_r$ with respect to the initial WIMP velocity. Since the WIMP velocity $v$ relative to the detector is of the order of the galactic rotation velocity $\sim 200$ km/s, the kinematics is non-relativistic. The expression for the nuclear recoil energy is easily found from energy and momentum conservation:

$$ E_r = \left( \frac{m_\chi}{2}\, v^2 \right) \times \frac{4\,m_N\,m_{\chi}}{(m_N+m_{\chi})^2} \times \cos^2\theta_r $$

\noindent Typical recoil energies are in the range 1-100 keV : for WIMP searches, this requires to use low threshold detectors, which are sensitive to individual energy deposits of this order of magnitude. To compute the WIMP-nucleon interaction rate, one needs the cross-section and the local phase-space density of WIMP:
\begin{itemize}
\item For a given momentum transfert $q$ we use the parametrization 

$$\frac{d\sigma}{dq^2} = \frac{\sigma_0}{4\,m_r^2\,v^2} F^2(q)$$

\noindent where $m_r$ is the reduced mass of the system, and $F(q)$ is a dimensionless ``form factor" such that $F(0)=1$. Since the maximum momentum transfert for a given $(v,m_r)$ is $q_{\rm max} = 2 v m_r$, the parameter $\sigma_0$ corresponds to the total cross-section in the case of $F(q)=1$.
\item We note $\rho_0$ the local WIMP mass density. The current observations contrain $\rho_0 \sim 0.3$ GeV/cm$^3$. The distribution of WIMP velocities relative to the terrestrial detector is noted $f_1(v)$.
\end{itemize}
The interaction rate per unit mass of detector for WIMPs in the velocity range $\left[v ; v+dv \right]$ is then given by:

\begin{equation}\label{eqn:diffrate}
dR = \left( \frac{\rho_0}{m_{\chi}\,m_N} \right) v\,\frac{d\sigma}{dq^2}\,f_1(v)\, dv\, dq^2
\end{equation}

\noindent After integration over the velocity distribution, this gives as a function of recoil energy $E_r=q^2/2m_N$:

$$ \frac{dR}{dE_r} = \frac{\sigma_0\,\rho_0}{2\,m_{\chi}\,m_r^2} \, F^2(q) \int_{v_{\rm min}}^{\infty} dv\, \frac{f_1(v)}{v}$$

\noindent where $v_{\rm min}=\sqrt{\frac{m_N\,E_r}{2\,m_r^2}}$. We use a maxwellian velocity distribution for the galactic WIMPs. Assuming that the detector is at rest with respect to the galactic halo, we have $f_1(v) \propto \times \frac{v^2}{v_0^3} e^{-v^2/v_0^2}$.  The integration is then straightforward and one finds

$$ \frac{dR}{dE_r} \propto \exp \left( -\frac{m_N\,E_r}{2\,m_r^2\,v_0^2}\right)$$

\noindent An approximately exponential recoil spectrum is therefore expected : as a consequence, no really precise spectral signature such as a peak may be used, and in addition most of the signal in a detector is expected at low recoil energies, which requires the energy threshold of all WIMP detectors to be well understood experimentally.

In fact, the Earth velocity with respect to the WIMP halo must be written as $v_e = v_0 \,(1.05 + 0.07 \cos \omega t)$ where $1.05\,v_0$ is the galactic velocity of the Sun and $\omega = 2\pi / 1\,$year. The $7\%$ modulation is due to the rotation of the Earth around the Sun. In the former calculation, $f_1(v)$ must be replaced by $f_1(|\vec{v}-\vec{v_e}|)$ and the instantaneous differential rate of WIMP interaction becomes

\begin{equation}\label{eqn:erfrate}
\frac{dR}{dE_r} = \frac{\sigma_0\,\rho_0}{4\,v_e\,m_{\chi}\,m_r^2}\, F^2(q) \left[ {\rm erf} \left( \frac{v_{\rm min}+v_e}{v_0} \right) - {\rm erf} \left(  \frac{v_{\rm min}-v_e}{v_0} \right) \right]
\end{equation}

In the conventional model which is currently used in order to measure or constrain cross-sections, and to compare experimental sensitivities, it is also assumed that the WIMP maxwellian distribution is truncated for velocities larger than the galactic escape velocity $v_{\rm esc} = 650$ km/s. This adds a small correction to the previous formula.

\subsection{WIMP-nucleus cross-section for neutralinos}

In this section we sketch qualitatively how the cross-section, ie. the parameters $\sigma_0$ and $F(q)$, is derived in a given particle physics model, and concentrate on the neutralino as an important example. We refer to~\cite{Jungman:1995df} for more details. Neutralinos interact with quarks through the exchange of Z bosons, squarks and higgses. In the same way as for the Fermi description of weak interactions, in the low-energy limit the neutralino-quark coupling may be written as a coupling between currents that are of scalar, vector, pseudo-scalar, axial or tensorial natures. Since the neutralino is a Majorana fermion, only axial and scalar couplings exist.

\subsubsection{Axial coupling : spin-dependent cross-section.}

The lagrangian for the axial neutralino-quark coupling, due to Z and squark exchange, is $\mathcal{L}_{q\chi} = d_q (\bar{\chi} \gamma^{\mu} \gamma_5 \chi)\,(\bar{q}\gamma_{\mu} \gamma_5 q)$ where the parameter $d_q$ is calculated for a given SUSY model. At the scale of a nucleon $n$ - proton or neutron - the matrix element  $\langle n | \bar{q}\gamma^{\mu} \gamma_5 | n \rangle$ is of the form $2\, S^{\mu} \Delta q^{(n)}$ where $S^{\mu}$ is the nucleon spin. The parameter $\Delta q^{(n)}$ depends on hadron physics and is estimated from lepton-proton diffusion experiments.
An effective neutralino-nucleon lagrangian is therefore obtained by summation over the relevant quark flavors. We define

$$ \sum_{q=u,d,s} 2\,d_q\,\Delta q^{(n)} \equiv 2 \sqrt{2} G_F \, a_{(n)} $$

\noindent With this convention, $a_p$ and $a_n$ are dimensionless parameters. The neutralino-nucleon coupling is then

$$\mathcal{L}_{n\chi}^{\rm eff} = 2 \sqrt{2} \,G_F \,a_{(n)}\, (\bar{\chi} \gamma^{\mu} \gamma_5 \chi)\,(\bar{n}\gamma_{\mu} \gamma_5 n)$$

\noindent The derivation of the neutralino-nucleus cross-section requires then some nuclear physics input which we won't detail here. For a nucleus of spin J, with $\langle S_p \rangle$ and $\langle S_n \rangle$ being the average spins ``carried" by protons and neutrons respectively, the cross-section at zero momentum transfert is

$$ \frac{d\sigma}{dq^2} (q=0) = \frac{8}{\pi v^2} G_F^2 \,\Lambda^2\, J(J+1) \;\;\;\;\; {\rm where} \;\;\;\;\; \Lambda = \frac{a_p \langle S_p \rangle + a_n \langle S_n \rangle}{J}$$

This is the so-called spin-dependent WIMP-nucleus cross section. There is an additionnal ``form factor" correction $S(q)$ to take into account in the case of non-zero momentum transfert. Quite often, it is a good approximation to say that the nucleus spin is carried mostly by one kind of nucleon. In that case, depending on the target nucleus, a measurement of $d\sigma / dq^2$ is almost a measurement of $a_p$ or $a_n$, or equivalently the proton and neutron spin-dependent cross-sections $\sigma_p^{\rm SD}$ and $\sigma_n^{\rm SD}$. Particularly appropriate nuclei for spin-dependent measurements are $^{73}$Ge ($S_p = 0$, $S_n=0.23$) and $^{131}$Xe ($S_p = -0.04$, $S_n=-0.24$) for $a_n$ measurements, and for example $^{19}$F ($S_p=0.46$, $S_n=0$) for $a_p$ measurements. The corresponding experiments are reviewed in section~\ref{sec:spinind}.

\subsubsection{Scalar coupling : spin-independent cross-section.}

Higgses and squark exchanges lead to a scalar neutralino-quark coupling $\mathcal{L}_{q\chi} = f_q (\bar{q}q)(\bar{\chi}\chi)$. To express the neutralino-nucleon coupling one needs the matrix element $m_q \langle n | \bar{q} q | n \rangle \equiv m_{(n)} f_{Tq}^{(n)}$, $(n)$ being a proton or neutron, and $q$ is a quark (u, d or s). The $f_{Tq}^{(n)}$ coefficients depend on hadron physics and are obtained from pion-nucleon diffusion measurements. In addition to coupling with quarks, there are also neutralino-gluon couplings which have an important contribution to the scalar nucleon coupling through QCD effects, which we do not describe here (see~\cite{Jungman:1995df}). The resulting neutralino-nucleon coupling is of the form

$$\mathcal{L}_{n\chi}^{\rm eff} = f_{(n)}\,(\bar{\chi}\chi)(\bar{n}n) \;\;\;\; {\rm where} \;\;\;\;\; f_{(n)} = \sum_{q=u,d,s} f_q \frac{f_{Tq}^{(n)}}{m_q} + \cdots$$

\noindent The additional gluon contributions to $f_p$ and $f_n$ are not explicited. It appears that the QCD terms and the coupling to the sea quark s are often dominant in the $f_{(n)}$ calculation. Therefore one has in general $f_p \simeq f_n$. For a given WIMP model, one can therefore define a single spin-independent WIMP-nucleon coupling, independent on the nature and spin of the nucleon. The summation over all nucleons of a nucleus is then simpler than for the spin-dependent case. The classical nuclear density form factor $F(q)$ (the Fourier transform of spatial density of nucleons in the nucleus) appears, and one gets, applying Fermi's golden rule:

$$ \frac{d\sigma}{dq^2} = [Z\,f_p + (A-Z)\,f_n]^2 \frac{F^2(q)}{\pi v^2}$$

\noindent We may then identify in this expression the $\sigma_0$ parameter defined above, and for $f_p = f_n$ one finds $\sigma_0(A)= \frac{A^2 m_r^2}{m_r^2(p)} \sigma_0$(nucleon). Therefore in the spin-independant channel, all experiments measure approximately the same parameter $\sigma_0$(nucleon) $\equiv \sigma^{\rm SI}$ independently of the target nucleus. Furthermore, contrarily to the spin-independant case, as long as the nuclear form factor $F(q)$ is of order unity (which is the case for low nuclear recoil energies), there is a coherence effect which boosts the WIMP-nucleus cross section by a factor $A^2 m_r^2(A)$. As a consequence, heavy nucleus targets are used in spin-independent direct searches, and at least for the most popular SUSY neutralino models, the spin-independent channel is currently the most sensitive to a potential WIMP signal.
We will therefore concentrate in the following sections on the experimental techniques that are mostly dedicated to this channel.

\subsection{Parameter uncertainties}

In the framework of MSSM, ``predictions" of the value of $\sigma^{\rm SI}$, $a_p$ and $a_n$ may be achieved by the use of exploratory scans in the cMSSM parameter space, which has 5 free parameters (or even a wider parameter space). These scans are strongly sensitive on the priors used for the parameter distribution (eg. flat or logarithmically distributed), as well as on existing contraints from collider experiments and the cosmological measurement of $\Omega_m$. The same parameter space benchmark points as for collider searches may also be used. For current direct detection experiments, a particularly interesting region of parameter space is the so-called focus point where the neutralino is higgsino-like : this favors Higgs exchanges in Feynman diagrams contributing to $\sigma^{\rm SI}$ and generates a relatively large WIMP-nucleon cross-section $\sigma^{\rm SI}\sim 10^{-8}$ pb.

While the uncertainty on the interaction rate for direct detection experiments is dominated by the unknown WIMP physics which governs the value of cross-sections, there are also other sources of errors due to more conventionnal physics used in the interaction rate calculation.
\begin{enumerate}
\item Uncertainties in hadronic and nuclear physics are still quite large, concerning for example the $f_{Ts}$ parameter or the spin content of individual nucleons in a given nucleus. They generate uncertainties of order unity on the WIMP-nucleus cross-section for a given WIMP model.
\item Uncertainties on the structure of our local dark matter halo. First, the truncated maxwellian model for the velocity distribution is a simple approximation, with two free parameters (its velocity dispersion related to the circular velocity at solar radius, and the escape velocity). The presence of streams could change the function $f_1(v)$. Second, the local value of dark matter density is infered from measurements of the Milky Way rotation curve and a modelization of its different components. While the canonical value is $\rho_0 = 0.3$ GeV/cm$^3$, there is certainly an uncertainty of a factor 2 or 3 (see for example~\cite{Catena:2009mf}). Furthermore other substructures may come in addition to the smooth component of the halo. If the Solar System were in a dark matter clump, the direct detection signal would be strongly boosted, but, from current simulations, this is unlikely at the Solar System radius. Perhaps more interestingly, some models of subhalo interactions within the Milky Way predict the existence of a dark matter disk in the same plane as the Milky Way disk~\cite{Read:2008fh}, with a local density $\rho_{\rm disk} \sim (0.25 - 1) \rho_{\rm halo}$. If such a disk exists, it could boost the expected direct detection signals.
\end{enumerate}

\subsection{Comparison with dark matter indirect detection}

In contrast to direct detection, indirect detection of WIMPs consists in observing fluxes of secondary particles created by WIMP annihilation in various astrophysical objects. The secondaries may be $\gamma$-rays, X-rays, neutrinos, or charged particles. Depending on the sources, the fluxes may be diffuse or point-like. In the most common models, there is in principle less undetermination in the cross-sections involved, which are directly related to the annihilation cross-section governing the relic density $\Omega_m$. On the other hand, astrophysical uncertainties related to these indirect detection channels are large : boost factors of several orders of magnitudes may arise due to the small-scale structure of the halo, and astrophysical backgrounds to the dark-matter related fluxes are difficult to control. As a consequence, the performances of direct and indirect detection channels may be compared but this comparison is in general model-dependant, and the systematics of both methods are different, making them largely complementary.

One indirect detection channel is particularly well-suited for comparison with the spin-dependent searches for proton-WIMP coupling ($\sigma_p^{\rm SD}$ measurements). Models show that the WIMP density clustered in the solar interior should be at equilibrium: the solar WIMP density is driven by the capture rate due to diffusion on the solar protons, which is itself proportional to the spin-dependant WIMP-proton cross-section $\sigma_p^{\rm SD}$. The annihilation of solar WIMPs generates, among others, a flux of high energy neutrinos in an energy range where no other solar neutrino background exists. Upper limits on the very high energy solar neutrino flux may therefore be translated into a limit on the spin-dependant WIMP-proton cross-section. Currently, the limits on $\sigma_p^{\rm SD}$ from Super-Kamiokande and Ice Cube~\cite{Abbasi:2009uz} are more constraining than limits obtained by direct detection.

\section{General principles of WIMP detection, first experiments}\label{sec:principle}

To measure WIMP-induced nuclear recoils, detectors with a massive target and a low detection threshold (typically a few keV) must be used. The expected signatures of WIMP interactions are the following:
\begin{itemize}
\item The interactions generate nuclear recoils, in contrast to electronic recoils from $\gamma$ radioactivity.
\item Due to the small interaction rate, only single interactions will be observed (no multiple interactions).
\item For the same reason, WIMP interactions are uniformly distributed in the detector volume, while interactions induced by external radioactivity, with penetration length smaller than the detector size, do concentrate near the surface of the detectors.
\item The recoil spectrum has an approximately exponential shape.
\item For the spin-independant channel, the interaction rate varies approximately as the square of the target nucleus : the use of several nuclear targets can serve as a first cross-check for a potential signal.
\item The movement of Earth in the galactic halo implies that 1) the distribution of recoil directions is not isotropic ; 2) the recoil rate has an annual modulation. These two signatures are studied in Sections~\ref{sec:direction} and~\ref{sec:modulation}.
\end{itemize}

The main challenge is the removal of environmental backgrounds due to radioactivity or cosmic-ray induced signals. The following backgrounds are of importance in the current WIMP searches~\cite{Formaggio:2004ge}:

\begin{enumerate}
\item Gamma-ray radioactivity, due to the intrinsic radioactivity of the surrounding materials as well as the detector itself. As an example, photomultiplier radioactivity is a major source of background for scintillating detectors. Another example is the presence of long-lived isotopes due to the cosmic activation of Argon ($^{39}$Ar with a 269-year lifetime) and Germanium ($^{65}$Zn, $^{68}$Ge with lifetimes of order a year), which generate an intrinsic $\gamma$ background for these nuclear targets. 

The intensity of radioactive backgrounds may be attenuated by several orders of magnitude by a careful selection of all materials in the experiment as well as by the use of shieldings (typically lead). In addition, since $\gamma$-rays generate electronic recoils, most detectors of current generation are designed to discriminate actively nuclear recoils from electronic recoils. Finally, the external gamma-ray radioactivity may be rejected using a position reconstruction which allows to define an inner, fiducial volume within a detector.
\item Beta radioactivity due to materials in the immediate neighbourhood of the detectors. The $\beta$ penetration depth in Germanium ranges from 350 nm at 10 keV to 700 $\mu$m at 1 MeV, so that $\beta$s generate surface interactions which require special discrimination for heat-and-ionization detectors, see Section~\ref{sec:heation}. A major source of $\beta$ radioactivity is the $^{210}$Pb, a daughter of radon present in the air with a 22-year lifetime.
\item Diffusion of fast neutrons. Fast neutrons are generated by interactions of cosmic-ray induced muons in the rocks and materials surrounding the detectors. They also originate from intrinseque radioactivity of the rock and surrounding materials, due to ($\alpha$,n) and fission reactions generated by U/Th traces. A neutron with kinetic energy of a few MeV generates a typical $\sim 10$ keV nuclear recoil after elastic diffusion. Therefore this recoil cannot be discriminated against WIMPs like $\gamma$-rays. To reduce this background, the experiments are located deep underground where the muon flux is attenuated. Polyethylen shields are also used to thermalize the fast neutrons coming from external sources, and hence slow them down before they reach the detectors. In addition, contrarily to WIMPs neutrons generate multiple interactions : multiplicity measurements are a way to reject this background.
\item The coherent diffusion of solar neutrinos in the detectors would completely mimic WIMP signals (single nuclear recoils). This background is expected to be relevant only for extremely large exposures achieved with at least ton-scale detectors~\cite{Monroe:2007xp}. The observation of this background would moreover constitute an interesting by-product of WIMP searches.
\end{enumerate}

Direct detection experiments belong to the category of ``rare event search experiments'', like experiments looking for proton disintegration or neutrinoless double beta searches. Since several decades, deep underground laboratories were built in order to host these experiments. As an example, the largest of them is the Gran Sasso laboratory~\cite{gransasso} located along a road tunnel under the Gran Sasso mountain in Italy. Some underground laboratories are located at the place of an ancient mine. Currently, several construction or extension projects are under way, such as the Ulisse extension of the Modane underground laboratory~\cite{lsm} in the Fr\'ejus tunnel between France and Italy, or the DUSEL facility~\cite{dusel} in the United States.

The idea of dark matter direct detection was elaborated in the 80s. In the end of 80s, first limits on the WIMP-nucleon cross-section as a function of WIMP mass were obtained using ultra-pure, semiconducting germanium detectors with an ionization measurement. As an example, the Oroville experiment~\cite{Caldwell:1988su}, initially conceived for double-beta search, was able to set limits excluding some WIMP models that were envisaged at that time, in particular a possible 4th generation heavy Dirac neutrino (LEP results on the number of neutrino flavors was not there yet). These ionization-only experiments were severely limited by their unability to reject the abundant background of $\gamma$-ray induced electronic recoils. Still, improved ionization-only high purity Ge detectors remain competitive for the low-mass WIMP search ($M_{\chi}\sim 5$ GeV), thanks to their remarkably low energy threshold. As an example, the CoGeNT collaboration~\cite{Aalseth:2008rx} instrumented a specific HPGe detector with an intrinsic electronic noise of $\sim 70$ eV, obtaining a sensitivity for $\sigma^{\rm SI}$ of $2\times 10^{-4}$ pb for $M_{\chi} = 8$ GeV.

In the beginning of the 90s, solid scintillators were developped for WIMP search: the target materials are NaI or CsI cristals, and scintillation signals are read with photomultipliers. The falltimes of the scintillation signals are different on average depending on whether the initial recoil in the crystal is nuclear or electronic. As a consequence, a statistical discrimination is possible between $\gamma$-induced events and potential WIMP interactions, based on the observed pulse shape distribution. This allowed an improvement in sensitivity with respect to Ge ionization detectors. Still, the impossibility to perform an event-by-event discrimination along with the presence of ``anomalous'' events eventually limited the spin-independent sensitivity of these detectors to $\sigma^{\rm SI}\sim 10^{-4}$ pb. Currently, solid scintillators remain competitive, first in spin-dependent searches (section~\ref{sec:spinind}), and second for annual modulation searches (section~\ref{sec:modulation}), since it is relatively easy to operate large masses of such detectors during long periods.

The best sensitivities in the spin-independent channel are currently obtained with two technologies, which we describe in the following sections: low temperature heat-and-ionization detectors and dual-phase noble liquids.

\section{Bolometric dark matter detectors}\label{sec:bolos}

The principle of these detectors is to measure, in addition to an ionization or scintillation signal, a ``heat" signal using a thermometer on the target to detect the temperature elevation due to a WIMP interaction.
For a given recoil energy, the ionization or scintillation yield provides a discriminating variable between electronic and nuclear recoils. This discrimination can be achieved on an event-by-event basis.

A cryogenic bolometer is made of an absorber : a crystal, with heat capacitance $C$, in contact with a thermal bath with a thermal leakage $G$ (expressed in J/K/sec). A thermometer measures the temperature variations of the absorber. For dark matter searches, bolometers are used in ``pulse" mode: after the interaction of an incident particle, a large fraction of its energy is converted into thermal energy. This instantaneous energy deposit $E_0$ in the absorber generates a temperature rise $\Delta T = E_0 / C$. The length of these temperature pulses is the time necessary for the cristal to thermalize with the thermal bath, of order $C/G$ : this is typically $\sim 100$ ms, therefore only small interaction rates may be detected (this is not a problem for WIMP searches). The bolometer sensitivity is inversely proportional to $C$. Since the Debye law $C(T) \sim T^3$ applies for semi-conductors, low energy depositions may be measured only at ultra-low temperatures, typically $T\sim 10 -100$ mK. Furthermore, the thermometer, consisting of a variable resistor $R(T)$ usually polarized at constant intensity, must have a temperature coefficient $\alpha = d \ln R / d \ln T$ as large as possible. Several technologies were developped, in particular:
\begin{itemize} 
\item Semi-conducting materials which have a negative $\alpha$, such as germanium cristals that are neutron transmutation doped (Ge-NTD).
\item Transition-Edge Sensors at the temperature of their supraconducting transition, with a large positive $\alpha$.
\end{itemize}
\noindent To reach stable temperatures of $\sim 20-40$~mK during continuous operation of several months, dilution cryostats must be used. This is a particularly strong technical constraint when large masses of detectors must be implemented, especially for bolometers, which require a low microphonic noise environment.

\subsection{Heat-and-ionization detectors}\label{sec:heation}

Heat-and-ionization bolometers are used by the EDELWEISS and CDMS collaborations. The absorber is a Germanium (or Silicium) cristal of typical mass 400 g nowadays, cooled down to 20 mK or 40 mK. Electrodes cover the cristal, and a voltage of order a few volts is applied in order to collect the ionization signals. No more than a few volts may be applied due to the low temperature. Also, with large voltages, the heat channel would only measure the Joule effect associated to charge drift, and would not add any information complementary to the ionization channel.

The ionization yield measurement results in a remarkable discrimination of $\gamma$-ray induced interactions within the absorber. An issue with these detectors comes with surface interaction discrimination: when an interaction takes place close to the electrode surface, the charge collection mechanism cannot be fully completed, resulting in a loss of ionization signal. A surface electronic recoil may therefore mimic a bulk nuclear recoil. A source of such surface events is the $\beta$ radioactivity due to $^{210}$Pb, which is a daughter element of radon. The ubiquity of radon in the air, together with the 22 year lifetime of $^{210}$Pb, make it difficult to reduce the presence of this background. Typical surface $\beta$ rates of a few events/kg/day are observed. These events of radioactive origin have limited EDELWEISS sensitivity to WIMP-nucleon cross-sections of $\sim 10^{-6}$ pb in 2003~\cite{Fiorucci:2006dx}.

The rejection of surface events by CDMS-II is currently based on their specific phonon measurement. The CDMS thermometer covers the surface of the absorber and is able to measure the athermal component of phonons generated during an interaction, which consists in a fast signal with respect to the thermal component. Surface interactions generate large athermal signals. A timing parameter using both the relative heat-ionization time and the heat pulse shape is estimated for each event and is used as a second discriminating variable, enabling the rejection of surface events at the price of a relative loss of efficiency.
In 2008, with 15 detectors in use, an effective exposure of 121 kg.days was unblinded, with a post-cut surface event background estimated to be 0.5 events. No WIMP candidate was observed above a recoil energy of 10 keV~\cite{Ahmed:2008eu}. In the end of 2009, the unblinding of 194 more kg.days lead to the observation of two low-energy WIMP interaction candidate events, for $0.8\pm0.3$ background events expected~\cite{Ahmed:2009zw}. Even with these two background events, an upper limit of $3.8\times 10^{-8}$ pb could be set on $\sigma^{\rm SI}$ for $m_{\chi} = 70$ GeV, which is the best sensitivity achieved so far for all direct detection searches.

The rejection of surface events for EDELWEISS-II is based on the recently developped InterDigit electrode design~\cite{Broniatowski:2009qu}. Plane electrodes are replaced by a set of concentric rings alternatively polarized for example at 4V and $-1.5$V on one side, and the opposite on the other side. This modifies the electric field near the detector surface. An interaction taking place in the bulk of the target generates charges only on the collecting (4V) electrodes, while a near-surface event also creates a charge signal on one of the field shaping (1.5V) electrodes. A $\beta$ rejection factor approaching $10^{-5}$ was demontrated using a $^{210}$Pb source.
With 10 germanium detectors, a first limit of $10^{-7}$ pb could be set after 6 months of data taking in 2009 with an effective exposure of 144 kg.d and a recoil energy threshold of 20 keV~\cite{Armengaud:2009hc}. One WIMP candidate was observed at 21 keV, with a preliminary background estimation of less than 0.23 events expected from neutron, beta and gamma interactions.

\subsection{Heat-and-scintillation detectors}\label{sec:heatscint}

In heat-and-scintillation bolometers, the absorber is a scintillating material instrumented with a thermometer. A second much lighter calorimeter faces the main absorber in order to detect the photon signal generated by an interaction in the target detector. The scintillation yield of each event is a discriminating variable between nuclear and electronic recoils.
An advantage of this technique is the large choice of materials available as scintillating absorbers.

The CRESST-II experiment used 10 bolometers with a CaWO$_4$ absorber to set a limit on $\sigma^{\rm SI}$ of $5\times 10^{-7}$ pb at 50 GeV. The sensitivity is limited by the presence of a background apparently due to the low scintillation yield for nuclear recoils in these detectors : WIMPs must be searched as heat-only interactions with no light signal. Such no-light  events are also due to other backgrounds related directly to the detector properties that are under study~\cite{Schmaler:2009bu}.

More ambitious project of bolometric dark matter detectors are planned both in Europe and in the USA. The CDMS program consists in a future installation of the Super-CDMS setup at the new deep SNOLab laboratory, with $\sim 100$ kg of detectors. A further ton-scale experiment could be installed at DUSEL. The european project EURECA consists in a multi-target array of cryogenic detectors installed in an extension of the LSM laboratory. These programs would require an unprecedented control of background noises as well as a heavy cryogenic infrastructure.

\section{Noble liquid detectors}\label{sec:noble}

At temperatures below 165 K and 88 K, Xenon and Argon behave as dense liquids, with good scintillation yields of $\sim 40\times 10^3$ photons/MeV, as well as reasonably good electron mobilities. An interaction in these liquids ionizes the medium and produces excited states of the Xenon or Argon atoms. The excited states in turn generate an UV luminescence signal. There are two excited states (singlet and triplet); for Argon, the triplet lifetime is $1.6 \,\mu$s, large enough to be measured and discriminated from the singlet state using a pulse shape analysis.

The principle of a dual-phase noble gaz TPC is the following : the Xenon (or Argon) is kept in a vessel in thermodynamical conditions where it is in equilibrium between liquid and gaz, the gaz phase being a small volume above the liquid phase. The vessel is instrumented with an array of detectors, typically photomultipliers. A strong, vertical electric field of order 1 kV/cm is applied within the whole volume. An interaction within the liquid phase generates a first direct scintillation signal, called $S1$. Depending on the ionizing power of the incoming particle, free electrons are created and drift vertically until they reach the gaz phase. When crossing the boundary between the two phases, a second light $S2$ is emitted due to the difference of amplitude of the electric fields in both phases. From the different signals of the PMT array, and their relative timings, the position of the primary interaction may be reconstructed. Furthermore, since the ionization yield is smaller for nuclear recoils, the $S2/S1$ ratio is used as a discriminating parameter between electronic and nuclear recoils.

\subsection{Xenon detectors}\label{sec:xenon}

The XENON10 experiment~\cite{Aprile:2010bt,Angle:2007uj} instrumented 22 kg of Xenon with 89 PMTs on the top and bottom sides of a TPC. The Xenon was kept at 180 K and 2.2 atm with a specifically developped pulse tube cryocooler. Calibrations show that the populations of electronic and nuclear recoil are not completely separated by the discriminating variable based on $S2/S1$: they overlap partly, therefore to reject most of the $\gamma$ background one must also reject $\sim 50\%$ of nuclear recoils, resulting in a $\sim 50\%$ WIMP search efficiency loss. A 5.4 kg fiducial volume was defined in the bulk of the TPC using the event position reconstruction. Since most of the radioactivity comes from surfaces (in particular PMTs), this allows an efficient self-shielding against the $\gamma$ background. In 2007 a WIMP search was accomplished using 136 kg.days of effective exposure in the energy range $4.5 < E_r < 27$ keV. Some background remained present especially for $E_r > 15$ keV. These ``anomalous" events are probably due to double interactions taking place in the detector, among which one takes place in a dead region of the liquid Xenon and generates $S1$ but no $S2$ signal. A limit on $\sigma^{\rm SI}$ of $4.4 \times 10^{-8}$ pb for $M_{\chi} = 30$  GeV was obtained, which is still the best published sensitivity for WIMPs with masses below 44 GeV. An important systematic uncertainty on this sensitivity is the exact value of liquid Xenon scintillation yield for low energy nuclear recoil~\cite{Manzur:2009hp}. Recent measurements show that the XENON-10 sensitivity is more probably $5.6 \times 10^{-8}$ pb for $M_{\chi}=30$ GeV.

Similar results were obtained by the ZEPLIN-III experiment in 2008~\cite{Lebedenko:2008gb}. More massive Xenon detectors are now under construction or in operation. In particular the XENON100 experiment~\cite{Aprile:2009yh} at the Gran Sasso laboratory is starting taking WIMP search data. It consists of a 170 kg Xenon TPC with a fiducial volume ranging from 30 to 50 kg depending on the position cuts. The overall radiopurity of the experiment was improved with respect to XENON10, and the side and bottom shields were equipped with PMTs to veto multiple scatterings. A more ambitious ``1 ton" phase is scheduled. A similar program is being developped in the US with the LUX experiment~\cite{Fiorucci:2009ak} consisting of a 350 kg Xe TPC. A particularity of LUX will be the use of a water shielding against $\gamma$ and neutron backgrounds.

A single-phase liquid Xenon program is also taking place in Japan at the Kamioka mine : the XMASS experiment consists of a 900 kg Xe sphere completely surrounded by low-radioactivity PMTs. Since only liquid Xenon is present and no electric field is applied, only the scintillation ($S1$) signal will be measured. The lack of $\gamma$-ray discrimination parameters should be compensated by the large Xe mass and the full PMT coverage with which a strict fiducial cut will be applied. 

\subsection{Argon detectors}\label{sec:argon}

Argon detectors work similarly to Xenon ones, with two main differences:
\begin{itemize}
\item The scintillation from triplet state may be discriminated from the singlet state thanks to a slow decay time. Since the singlet/triplet ratio depends on the ionizing power of the incident particle, the pulse shape analysis of $S1$ signals provides a second, independent discriminating variable to reject electron recoils.
\item On the other hand Argon contains naturally the $^{39}$Ar radioactive isotope generating a large electron recoil background in the fiducial volume of a detector. Argon reservoirs with a low $^{39}$Ar concentration have been found in underground cavities and may be used in the future. Methods of depletion in $^{39}$Ar are also investigated.
\end{itemize}

The WArP experiment~\cite{Benetti:2007cd} at the Gran Sasso laboratory has demonstrated the efficiency of the combination of $S2/S1$ + pulse-shape discriminations to reject electron recoils with a quoted rejection factor of $\sim 10^{-8}$. A first WIMP search with a 2.6 kg prototype lead to a sensitivity of $\sim 10^{-6}$ pb. This sensitivity was limited by a large $\gamma$ background and a small number of instrumented PMTs. A 140 kg dual-phase Argon TPC is now being installed at Gran Sasso, with a large coverage of low-radioactivity PMTs and an active veto. We only mention here the existence of other large mass projects for either dual-phase (ArDM) or single-phase (DEAP/CLEAN) argon detectors, that are also being planned or commissionned.

\subsection{Comparison with bolometers}
At present, both noble liquids and germanium bolometers are the most competitive techniques to probe the spin-independent channel in direct detection of WIMP. Both technologies require heavy cryogenic set-ups, the cryogeny for Ge bolometers being even more imposing for an equivalent target mass due to the necessity to cool-down massive targets at 20-40 mK. The definition of a fiducial volume may now be done for both noble liquids and bolometers. For noble liquids, the self-shielding effect will be more and more efficient as the instrumented mass grows. For bolometers, since the individual bolometer mass will be limited for heat capacitance reasons, the detector segmentation will be important. The electronic recoil vs nuclear recoil discrimination is much less efficient for Xenon detectors than for bolometers. This is less the case for Argon TPCs, but current Argon detectors must face a stronger internal $\gamma$-ray background. While it seems that going to larger-mass detectors will be easier with noble liquid detectors, there is a real complementarity between these techniques which would allow for an efficient cross-check in case of detection of a WIMP signal.

\section{Complementary detection channels}\label{sec:complementary}

Until now we focussed on the search for a spin-independent WIMP-nucleon coupling by measuring a nuclear recoil spectrum. There are other direct detection channels, which are complementary in the sense that a first spin-independent direct detection signal could be confirmed by one of these channels. For the most standard WIMP models, and with current technologies, these channels are less sensitive than the former one, but they have advantages of their own.

\subsection{Spin-dependent WIMP-nucleon cross-section}\label{sec:spinind}

As described above, this component of the WIMP-nucleon interaction cross-section depends strongly on the spin content of the target nucleus. Measurements or limits on the WIMP interaction rate on a specific nuclear target result in a specific constraint on the WIMP coupling coefficients $a_p$ and $a_n$, or equivalently $\sigma_p^{\rm SD}$ and $\sigma_n^{\rm SD}$. We provide here only a short selection of measurements, which are currently the most competitive.

\begin{itemize}
\item The WIMP-neutron spin-dependent cross-section is currently best constrained by the XENON10 experiment, using the same data as described in Section~\ref{sec:xenon}. The constraint on $a_n$ makes use of the $\sim 20\%$ natural abundance of the $^{129}$Xe and $^{131}$Xe isotopes. A limit of $5\times 10^{-3}$ pb was set on $\sigma_{n}^{\rm SD}$ for $M_{\chi} = 30$ GeV~\cite{Angle:2008we}.
\item Solid scintillators like NaI and CsI provide competitive constraints on $a_p$. As an example, the KIMS experiment~\cite{Lee.:2007qn} in Korea using CsI detectors has obtained in 2007 the best sensitivity on $a_p$ for high-mass WIMPs. As explained above, these detectors do not provide an event-by-event discrimination of the gamma background, but both the high exposure achieved (3400 kg.days) and the use of a statistical pulse-shape discrimination using the falltime of signals enabled to reach a sensitivity of $\sim 0.2$ pb for $\sigma_p^{SD}$.
\item The superheated droplet technique, used by the PICASSO~\cite{Archambault:2009sm} and COUPP~\cite{Behnke:2008zza} experiments currently provides the best sensitivity to $a_p$ at low WIMP mass. The $^{19}$F is used as a target in, for example, liquid CF$_3$I for COUPP. In appropriate temperature and pressure conditions, the interaction of a particle in the detector creates a bubble in the target volume, only for nuclear recoils. This efficiently suppresses the $\gamma$ background. Temperature or pressure scans enable to constrain the energy distribution of the observed nuclear recoils in a WIMP search. A sensitivity of 0.16 pb on $\sigma_p^{SD}$ for a WIMP mass of 24 GeV was recently reached by PICASSO.
\end{itemize}

\subsection{Directionnality of nuclear recoils}\label{sec:direction}

Due to the average relative velocity of the Solar System with respect to the dark matter halo, a strong anisotropy is expected in the WIMP-induced nuclear recoil distribution in the direction opposite to the Solar System velocity in the halo~\cite{Spergel:1987kx}. Using the notations of section~\ref{sec:theory}, and $\phi$ being the angle of the nuclear recoil with respect to the direction $(\ell,b)=(90^{\circ},0^{\circ})$ in galactic coordinates (which is approximately located in the Cygnus region, hence the expression ``Cygnus wind"), the recoil angular distribution is:

$$ \frac{dR}{dE_r\,d\cos \phi} \propto \exp \left( - \frac{(v_e \cos \phi - v_{\rm min})^2}{v_0^2} \right)$$

\noindent Since the Solar System velocity is of the same order of magnitude than $v_0$, the modulation amplitude is large, of order $\sim 80\%$ at appropriate recoil energies. Observing such a modulation in a population of nuclear recoils would provide a strong evidence that these recoils are indeed induced by WIMPs. To measure the direction of low-energy nuclear recoils in a WIMP-search experiment, a favored technique is the use of gazeous detectors. The corresponding challenge is the necessity to build very large volumes of detectors, with a large number of associated readout channels. Several R\&D developments were made in the past years to accomplish this~\cite{Ahlen:2009ev}. Let us mention two recent achievements as examples of the current activity:
\begin{itemize}
\item The DM-TPC collaboration, using a CF$_4$ gazeous TPC instrumented with a CCD camera, has observed the so-called head-tail effect for energies down to 100 keV. The asymetry in observed traces allowed the reconstruction of the vector direction of the recoils.
\item The MIMAC project has recently demonstrated the ability to reconstruct low energy traces at $\sim 6$ keV using micromegas detectors in a gazeous $^3$He TPC.
\end{itemize}

\subsection{Annual modulation of the recoil rate}\label{sec:modulation}

The Earth rotation around the Sun at $\sim 30$ km/s generates an annual modulation of the Earth-WIMP gaz velocity $v_e$, and therefore an annual variation of the shape and amplitude of the WIMP-induced  nuclear recoil spectrum, according to Eq.~\ref{eqn:erfrate}. At a given recoil energy, the interaction rate is sinusoidally modulated. The amplitude of the modulation depends on the recoil energy and is typically $\sim 7\%$. The observation of such a modulation in a nuclear recoil spectrum therefore constitutes one more test of the WIMP hypothesis. Since the modulation is weak, a high statistics of WIMP-induced nuclear recoils as well as a good control of long-term detector stability are required in order to perform such a test.

Since 1996, the DAMA experiment uses NaI(Tl) solid scintillators in order to search for such a modulation. With these cristals, an exceptionnally high exposure of 0.82 ton.year could be collected, using first 100 kg during 7 years, and then 250 kg during 4 years in the DAMA/LIBRA setup~\cite{Bernabei:2008yh,Bernabei:2008yi}. As explained above, the $\gamma$-ray background cannot be rejected on an event-by-event basis in these scintillators, and actually the DAMA collaboration does not perform any pulse shape rejection of electronic recoils in its annual modulation search. Therefore an integral spectrum, including radioactivity backgrounds as well as possible WIMP interactions, is observed, and an annual modulation signal is searched for in this spectrum for energies $E>2$ keV. The combination of all DAMA/NaI and DAMA/LIBRA data shows a statistically compelling ($8\,\sigma$) modulation signal at low energies, for $2 \lesssim E \lesssim 4$ keV. This modulation signal is compatible with a WIMP interpretation: 
\begin{itemize}
\item Making use of the segmentation in different cristals, the modulation is only observed for single events.
\item The dependance of the modulation with energy as well as the observed phase of this modulation, with a maximum in the beginning of June, are compatible with WIMP predictions.
\end{itemize}

\noindent Still, this result remains controversial from an experimental point of view. In particular, the modulated signal is near 3 keV, an energy where a peak at 3.2 keV associated to $^{40}$K dominates the radioactivity background, and where the detector efficiency drops sharply, opening the way to several possible systematic effects. The KIMS experiment (see section~\ref{sec:spinind}), now also searching for an annual modulation in a different underground laboratory and with different cristals (CsI), could provide a possible direct cross-check of the modulation signal.

From a phenomenological point of view, the DAMA signal is merely compatible with other searches, both in the spin-independent and spin-dependent channels. An interpretation of DAMA signal as due to a spin-dependent WIMP-proton coupling is now severely constrained by KIMS, COUPP and PICASSO. All currently competitive bolometric and noble liquid searches exclude the DAMA signal in the hypothesis of a purely spin-independent coupling, except for very-low mass WIMPs $M_{\chi} < 10$ GeV. In this low-mass range, the CoGeNT experiment (see section~\ref{sec:principle}) could put constraints on the DAMA signal, but a ``window'' remains clearly open, mostly due to the uncertainties on detector properties (such as a hypothetical channeling effect in NaI cristals) and on astrophysical parameters (especially the local WIMP velocity distribution).

An abundant litterature therefore exists in order to interpret the potential DAMA signal in light of other null-result searches. As an example we focus here on a plausible phenomenological explanation which has been the subject of recent interest, named inelastic dark matter~\cite{TuckerSmith:2001hy} : when scattering  on nuclei, WIMPs $\chi$ could be able to jump to an excited state $\chi^*$ with $m_{\chi^*} = m_{\chi} + \delta$, $\delta \sim 100$ keV. This induces a change of kinematics with respect to the elastic scattering, suppressing the recoil spectrum for low energies and increasing the relative annual modulation signal. The new free parameter $\delta$ opens more possibilities but still the analysis or re-analysis of null-search results, especially of XENON10 and CDMS, is able to put severe constraints on this scenario~\cite{Angle:2009xb,Ahmed:2009zw}.

\section{Direct detection of axion dark matter}\label{sec:axions}

Axions or axion-like particles are coupled to the photon field through the interaction term $\mathcal{L}_{a\gamma} \sim g_{a\gamma\gamma} F \tilde{F} a = g_{a\gamma\gamma} (\vec{E}\cdot\vec{B})\,a$. The coupling constant $g_{a\gamma\gamma}$ is not strongly constrained for generic models, but in the case of the standard Peccei-Quinn axion models, $g_{a\gamma\gamma} \propto m_a$.

The mass of dark matter axions is constrained by cosmological and astrophysical observations. As dark matter axions must be non-thermal relics of the Big Bang, predictions are more model-dependant than in the case of thermal relics like WIMPs. In general, $m_a \leq 10^{-6}$ eV would lead to a relic density $\Omega \geq 1$. For $m_a \geq 1$ eV, relic axions would constitute hot dark matter.

Therefore, in the specific framework of Peccei-Quinn axions with a relic density $\Omega_M = \Omega_a \sim 0.3$, a region of particular interest is $10^{-6} \leq m_a \lesssim 10^{-3}$ eV, and the corresponding coupling $g_{a\gamma\gamma} \sim 10^{-17} - 10^{-13}$ GeV$^{-1}$.

Current and proposed experiments using resonant cavities are exploring this region of parameter space. Dark matter axions are searched by looking for the conversion of a magnetic field into an electric field in a resonant microwave cavity.

As an example, in the case of the ADMX experiment~\cite{Duffy:2006aa}, the resonant cavity is filled with an 8T magnetic field, and the electric field is measured with an appropriate low-noise readout, now using a SQUID amplifier. The electric field power is measured for different values of the resonance frequency $f$ of the cavity which is varied mechanically by changing its effective dimensions. The signal from an axion with mass $m_a$ would appear as a peak in the electric field intensity vs $f$ for $f=m_a$. At a given mass, the sensitivity to $g_{a\gamma\gamma}$ is related to the measured electric field noise, which is limited by the readout noise as well as the thermal noise of the cavity. Current ADMX results already explore the Peccei-Quinn axion models in a narrow mass range $1.9\times 10^{-6} < m_a < 3.5 \times 10^{-6}$ eV. Upcoming upgrades as well as other projects such as CARRACK aim at extending the range of mass scan as well as the sensitivity to $g_{a\gamma\gamma}$.

\section*{Acknowledgments}
I thank the Ecole de Gif organizers for having provided such a nice atmosphere during the school.

\end{document}